\documentclass[12pt,a4paper]{amsart}

\newcommand{\CC}{\mathbb{C}}
\newcommand{\RR}{\mathbb{R}}

\newcommand{\NN}{\mathbb{N}}

\newcommand{\VV}{\mathcal{V}}
\newcommand{\UU}{\mathcal{U}}

\newcommand{\spec}{\mathop{\mathrm{spec}}\nolimits}
\newcommand{\diag}{\mathop{\mathrm{diag}}}
\newcommand{\ess}{\mathrm{ess}}

\newtheorem{theorem}{Theorem}
\newtheorem{prop}[theorem]{Proposition}

\newtheorem{corol}[theorem]{Corollary}

\begin{document}

\title[Hamiltonians with degenerate bottom]{Variational principle for Hamiltonians with degenerate bottom}

\author{Konstantin Pankrashkin}

\address{  D\'epartement de Math\'ematiques, Universit\'e Paris Sud,
  91405 Orsay, France \&
  Institut
  f{\"u}r Mathematik, Humboldt-Universit{\"a}t zu Berlin,  10099 Berlin, Germany }
  
\email{const@math.hu-berlin.de}



\date{\today}

\begin{abstract}
 We consider perturbations of Hamiltonians whose Fourier symbol attains its minimum
along a hypersurface. Such operators arise  in several domains, like spintronics,
theory of supercondictivity, or theory of superfluidity. 
Variational estimates for the number of eigenvalues
below the essential spectrum in terms of the perturbation potential are provided.
In particular, we provide an elementary proof that
negative potentials lead to an infinite discrete spectrum.
\end{abstract}

\maketitle

We are studying quantum Hamiltonians $H=H_0 +V$ acting on $L^2(\RR^n)$, $n\ge 2$,
where $V$ is a potential
and $H_0$ is a self-adjoint (pseudodifferential) operator whose Fourier
symbol $H_0(p)$ attains its minimal value
on a certain $(n-1)$-dimensional submanifold $\Gamma$
of $\RR^n$ (surface of extrema). A possible example for $H_0$ is 
the Hamiltonian
\begin{equation}
         \label{eq-ex1}
H_0(p)=\Delta+\dfrac{\big(|p|-p_0\big)^2}{2\mu}, \quad \Delta,\mu,p_0>0, \quad p\in\RR^3,
\end{equation}
arising in the study of the roton spectrum in liquid helium II \cite{KC}
and introduced by Landau \cite{La}. Another example can be
the three-dimensional Hamiltonian
\begin{equation}
         \label{eq-ex2}
H_0(p)=(p^2-\mu)\dfrac{e^{\beta(p^2-\mu)}+1}{e^{\beta(p^2-\mu)}-1}, \quad \mu,\beta>0
\end{equation}
which arised recently in the theory of supercondictivity \cite{bcs1,bcs2}; we refer
to the papers cited for the physical meaning of all the constants.
Similar operators appear in the study
of matrix Hamiltonians related to the spintronics (see below)
and in the elasticity theory \cite{CF}. A class of operators of the above type
were studied in \cite{LSW} using the Birman-Schwinger approach.
In this paper we are going to provide variational estimates
for $H_0+V$ with localized potentials $V$ in a possibly simplest form.
In particular,
we provide an elementary proof for the existence of infinitely many eigenvalues below the essential
spectrum for perturbations by negative potentials.
Our estimates can be viewed as a generalization of the classical
result: the existence of a negative eigenvalue for $-\Delta+V$
in dimensions one and two is guranteed by the condition $\int V(x)dx<0$.

It seems that the presence of an infinite discrete spectrum in the physics 
literature in such a setting
has been observed first \cite{CM} on example of rotationally invariant perturbations of the Rashba Hamiltonian. 
In the joint papers \cite{BGP1,BGP2} we gave a rigorous justification
for a class of spin-orbit Hamiltonians and rather general potentials, including interactions
supported by null sets. The proof was variational and used explicitly
the specific properties of two-dimensional systems.
Here we develop this idea in a different direction
and use the one-dimensional character of the dynamics in the direction transversal
to the surface of extrema.

Let us list our assumptions. Below we consider
a self-adjoint operator $H_0=H_0(-i\nabla)$, where
$\RR^n\ni p\mapsto H_0(p)\in\RR$ is a semibounded below continuous function attaining
its minimum value $\min H_0=m$.
Denote $\Gamma=\{p\in\RR^n:\,H_0(p)=m\}$; we will assume that for some
domain $\Omega\subset\RR^n$ the intersection $S=\Omega\cap\Gamma$ is
a smooth $(n-1)$-dimensional submanifold of $\RR^n$;
by $\omega$ we denote the induced volume form on $S$. 
Without loss of generality we assume that $\overline S$ is compact and orientable
(otherwise one can take a smaller $\Omega$).
We also suppose that $H_0$ is at least of $C^2$ class near $S$.

For both the Hamiltonians \eqref{eq-ex1} and \eqref{eq-ex2} one take $\Omega=\RR^3$.
For \eqref{eq-ex1}, one has $m=0$ and $S$ is the sphere of radius $p_0$
centered at the origin. For \eqref{eq-ex1} one has $m=2\beta^{-1}$
and $S$ is the sphere of radius $\sqrt{\mu}$ centered at the origin.

Consider a real-valued potential $V\in L^1(\RR^n)$.
We wil assume that the operator $H=H_0+V$ defined as a form sum
is self-adjoint with
\begin{equation}
        \label{eq-comp}
\inf\spec_{\ess}(H_0+V)=\inf\spec_{\ess}H_0=m. 
\end{equation}
For both the Hamiltonians \eqref{eq-ex1} and \eqref{eq-ex2} 
the assumption \eqref{eq-comp} holds for $V\in L^{3/2}(\RR^3)\cap L^1(\RR^3)$;
indeed, such $V$ is relatively compact with respect to the Laplacian.
As the difference $(H_0-\text{Laplacian})$ is infintely small with respect to
the Laplacian, $V$ is a relatively compact
perturbation of $H_0$ as well.

In what follows we will work in the $p$-representation.
The operator $H$ is then associated with the bilinear form
\[
\langle f, H f\rangle=\int_{\RR^n} H_0(p)|f(p)|^2d+\int_{\RR^n}\int_{\RR^n} \Hat V(p-p') \overline{f(p)}f(p') dp\,dp',
\]
where $\Hat V$ is the Fourier transform of $V$; in our case $\Hat V$
is a bounded continuous function due to $V\in L^1(\RR^n)$.
By $\VV$ we denote the operator on $L^2(S,\omega)$ acting by the rule
\[
\VV f(s)=\int_S \Hat V(s-s') f(s') \omega(ds').
\]
We note that such an operator already appeared in \cite{LSW}.

\begin{prop}\label{prop1}
The number of eigenvalues of $H$ below $m$ is not less than
the number of negative eigenavlues for $\VV$ counting multiplicities.
\end{prop}

\begin{proof}

Let $n(s)$ be a unit normal vector to $S$ at a point $s\in S$ and depend on $s$
continuously.
For $r>0$ consider the map $L: S\times(-r,r)\to \RR^n$,
$(s,t)\mapsto s+ tn(s)$; we choose $r$ sufficiently
small in order that $L$ becomes a diffeomorphism between $S\times(-r,r)$
and $L\big(S\times(-r,r)\big)$.
Note that due to the above assumption on $H_0$ and $S$
one has $H\big(L(s,t)\big)-m\le C t^2$ for $t\to0$ with some $C>0$.

Consider two arbitrary function $\Psi_1,\Psi_2\in L^2(S,\omega)$.
Take an arbitrary function $\varphi\in C_0^\infty(\RR)$ with $\int \varphi=1$ and
$\varepsilon>0$. Consider functions $f_j^\varepsilon\in L^2(\RR^n)$
given by
\begin{equation}
             \label{eq-fj}
f_j^\varepsilon(p)=\begin{cases}
\dfrac{1}{\varepsilon}\, \varphi\Big(\dfrac{t}{\varepsilon}\Big)
\Psi_j(s), &  p=L(s,t),\quad (s,t)\in S\times(-r,r),\\
0 & \text{otherwise.}
\end{cases}
\end{equation}
Clearly,
\begin{multline*}
\langle f^\varepsilon_1, (H-m) f^\varepsilon_2\rangle=
\int_{\RR^n} \overline{f^\varepsilon_1(p)} \big(H_0(p)-m\big) f^\varepsilon_2(p) dp\\
+
\int_{\RR^n} \int_{\RR^n} \Hat V(p-p')  \overline{f^\varepsilon_1(p)}  f^\varepsilon_2(p')\, dp\, dp'.
\end{multline*}
One has $dp=\rho(s,t)\omega(ds) dt$ with $\rho(s,t)=1+O(t)$ for $t\to0$
uniformly in $s\in S$, hence
\begin{multline*}
\left|\int_{\RR^n} (H_0(p)-m) \overline{f^\varepsilon_1(p)}  f^\varepsilon_1(p) dp\right|\\
=\left|\dfrac{1}{\varepsilon^2 }\int_{-r}^r \int_S
(H\big(L(s,t)\big)-m) \Big|\varphi\Big(\dfrac{t}{\varepsilon}\Big)\Big|^2 \overline{\Psi_1(s)} \Psi_2(s)
\rho(s,t)\omega(ds)dt 
\right|\\
\le 
C\left|\dfrac{1}{\varepsilon^2 }\int_{-r}^r \int_S
t^2 \Big|\varphi\Big(\dfrac{t}{\varepsilon}\Big)\Big|^2 
\overline{\Psi_1(s)} \Psi_2(s)
\rho(s,t)\omega(ds)dt 
\right|\\
\le 
C\varepsilon\left|\int_{-r/\varepsilon}^{r/\varepsilon} \int_S
t^2 \big|\varphi(t)\big|^2 \overline{\Psi_1(s)} \Psi_2(s)
\rho(s,\varepsilon t)\omega(ds)dt 
\right|=O(\varepsilon).
\end{multline*}
On the other hand, for any bounded continuous function $v:\RR^n\times\RR^n\to\RR$ one has
\begin{multline}
            \label{eq-v1}
\int_{\RR^n} \int_{\RR^n} v(p,p')  \overline{f^\varepsilon_1(p)}  f^\varepsilon_2(p')\, dp\, dp'\\
=
\dfrac{1}{\varepsilon^2 }\int_{-r}^r \int_{-r}^r \int_S \int_S
v \big(L(s,t),L(s',t')\big) \overline{\varphi\Big(\dfrac{t}{\varepsilon}\Big)} \varphi\Big(\dfrac{t'}{\varepsilon}\Big)\\
\times 
\overline{\Psi_1(s)} \Psi_2(s' ) \rho(s,t) \rho(s',t')\,
\omega(ds) \omega(ds') dt\, dt'\\
\int_{-r/\varepsilon}^{r/\varepsilon} \int_{-r/\varepsilon}^{r/\varepsilon} \int_S \int_S
v \big(L(s,\varepsilon t),L(s',\varepsilon t')\big) \overline{\varphi(t)} \varphi(t')\\
\times \overline{\Psi_1(s)} \Psi_2(s' ) \rho(s,\varepsilon t) \rho(s',\varepsilon t')\,
\omega(ds) \omega(ds') dt\, dt'=:I(\varepsilon).
\end{multline}
Due to the obvious estimate
\begin{multline*}
\bigg|\int_S \int_S
v \big(L(s,\varepsilon t),L(s',\varepsilon t')\big)\\
\times  \overline{\Psi_1(s)} \Psi_2(s' ) \rho(s,\varepsilon t) \rho(s',\varepsilon t')\,
\omega(ds) \omega(ds')\bigg|\\
\le \Tilde C \int_S  \big|\Psi_1(s)\big| \omega(ds)\, \int_S   |\Psi_2(s)| \omega(ds)
\end{multline*}
with $\Tilde C=\sup_{p,p'\in\RR^n} |v(p,p')| \sup_{(s,t)\in S\times(-r,r)} |\rho(s,t)|$,
one has, by the Lebesgue dominated convergence,
\begin{multline}
         \label{eq-v2}
\lim_{\varepsilon\to 0} I(\varepsilon)
=
\int_S \int_S
v \big(L(s,0),L(s',0)\big)  \overline{\Psi_1(s)} \Psi_2(s' ) \omega(ds) \omega(ds')\\
=  \int_S \int_S
v(s,s')  \overline{\Psi_1(s)} \Psi_2(s' ) \omega(ds) \omega(ds').
\end{multline}
Taking $v(p,p')=\Hat V(p-p')$, we have shown that for any $\Psi_1,\Psi_2\in L^2(S,\omega)$
and the functions $f^\varepsilon_1$, $f^\varepsilon_2$
given by \eqref{eq-fj} one has
\[
\lim_{\varepsilon\to 0}
\langle f^\varepsilon_1, (H-m)f^\varepsilon_2\rangle
=\langle \Psi_1,\VV \Psi_2\rangle.
\]

Assume now that $\VV$ has $N$ negative eigenvalues $E_1,\dots,E_N$
and let $\Psi_1,\dots,\Psi_N$ be the corresponding normalized
eigenfunctions orthogonal to each other. Consider
the functions $f^\varepsilon_j$, $j=1,\dots,N$, given by \eqref{eq-fj}.
Then, by the above arguments, the matrix 
$h(\varepsilon)=\big(\langle f^\varepsilon_j, (H-m) f^\varepsilon_k\rangle\big)$ converges to $\diag(E_1,\dots,E_N)$.
In particular, $h(\varepsilon)$ is negative definite for
suffiently small $\varepsilon$, which means, by the variational principle,
 that $H$ has at least $N$ eigenvalues below $m$.
\end{proof}
Due to the obvious estimate
\[
\int_S \int_S |\Hat V(s-s')|^2 \omega(ds)\,\omega(ds') <\infty
\]
$\VV$ is a Hilbert-Schmidt operator and hence compact, which
implies $\spec_{\ess} \VV=\{0\}$. 

\begin{prop}\label{prop2}
If $V\le 0$ and $V\not\equiv 0$, then the discrete spectrum of $\VV$
consists of an infinite sequence of negative eigenvalues converging to $0$,
and $0$ is not an eigenvalue.
\end{prop}

\begin{proof}
Let $f\in L^2(S,\omega)$.
One has
\begin{multline*}
\langle f,\VV f\rangle = \int_S \int_S \Hat V(s-s') \overline{f(s)} f(s') \omega(ds) \omega(ds')\\
=
\dfrac{1}{(2\pi)^{n/2}}\int _S \int_S \int_\RR V(x) e^{i\langle s'-s,x\rangle} \overline{f(s)} f(s') dx\,\omega(ds)\, \omega(ds')\\
=\dfrac{1}{(2\pi)^{n/2}}\int_\RR V(x) \big| g(x)\big|^2\,dx\le 0
\end{multline*}
with
\begin{equation}
           \label{eq-gf}
g(x):=\int_S f(s) e^{-i\langle s, x\rangle} \omega(ds).
\end{equation}
Therefore, $\spec \VV\subset (-\infty,0]$.

Assume that $\langle f,\VV f\rangle=0$ for some $f$.
The function $g$ in \eqref{eq-gf} is analytic as
the Fourier transform of the compactly supported distribution
$(2\pi)^{n/2} f(s)\delta_S(s)$, where $\delta_S$ is the delta measure concentrated on $S$.
To have $\langle f,\VV f\rangle=0$ the function $g$ must vanish on a set of non-zero Lebesgue measure
(the support of $V$) and hence vanish everywhere. As the Fourier transform is a bijection on the set of tempered distributions,
this means $f=0$. Therefore, $0$ cannot be an eigenvalue of $\VV$, and it remains to recall that $\VV$ is compact.
\end{proof}

Combining propositions \ref{prop1} and \ref{prop2}
one arrives at
\begin{corol}\label{corol3}
If $V\le 0$ and $V\not\equiv 0$, then $H$ has infinitely many eigenvalues below
the essential spectrum.
\end{corol}

If the condition $V\le 0$ does not hold, one still can try to estimate
the number of negative eigenvalues for $\VV$ using the values of the Fourier transform at some points. Due to  $\spec_{\ess} \VV= \{0\}$
the number of negative eigenvalues for $\VV$ can be estimated
using the variational principle as well.

\begin{prop}\label{prop4}
Let $N\in \NN$.
Assume that there exist points $s_j\in S$, $j=1,\dots,N$, such that the matrix
$\big(\Hat V(s_j-s_k)\big)$ is negative definite, then $V$
has at least $N$ negative eigenvalues and hence $H$ has at least
$N$ eigenvalues below $m$.
\end{prop}

\begin{proof}
Fix some neighborhoods $S_j\subset S$ of $s_j$ such that
there exists diffeomorphisms $J_j: B\to S_j$, where
$B$ is the unit ball centered at the origin in $\RR^{n-1}$.
Without loss of generality we assume $J_j(0)=s_j$.
Let us take a function $\varphi\in C^\infty_0(\RR^{n-1})$
with $D_j(0)\int \varphi=1$, where $D_j$ is the Jacobian for $J_j$.
Denote $\Psi^\varepsilon_j(s)=\varepsilon^{1-n}\varphi(\varepsilon^{-1}J^{-1}_j(s))\chi_{B}(J_j^{-1}(s))$;
clearly, $\Psi^\varepsilon_j\in L^2(S,\omega)$.
One has
\begin{multline*}
\langle \Psi^\varepsilon_j, \VV \Psi^\varepsilon_k\rangle=\int_S \int_S \overline{\Psi^\varepsilon_j(s)} \Hat V(s-s') \Psi^\varepsilon_k(s')
\omega(ds)\,\omega(ds')\\
=\varepsilon^{2-2n}
\int_B \int_B \overline{\varphi(u/\varepsilon)} \Hat V(J_j(u)-J_k(u')) \varphi(u'/\varepsilon)
D_j(u) D_k(u') du\,du'\\
\int_{B/\varepsilon} \int_{B/\varepsilon} \overline{\varphi(u)} \Hat V(J_j(\varepsilon u)-J_k(\varepsilon u')) \varphi(u')
D_j(\varepsilon u) D_k(\varepsilon u') du\,du'\\
\stackrel{\varepsilon\to 0}{\longrightarrow}
\Hat V(s_j-s_k).
\end{multline*}
Therefore, the matrix $\big(\langle \Psi^\varepsilon_j, \VV \Psi^\varepsilon_k\rangle\big)$
is negative defnite for small $\varepsilon$.
The rest follows from the variational principle and proposition \ref{prop1}. 
\end{proof}

Taking $N=1$ in proposition~\ref{prop4} we obtain a simple condition resembling that for
perturbations of the Laplacian in one and two dimensions.
\begin{corol}\label{corol5}
If $\int V(x)dx<0$, then $H$ has at least one eigenvalue below $m$.
\end{corol}

We note that corollary~\ref{corol3} can be also obtained from proposition~\ref{prop4}
because for $V\le 0$  and $V\not\equiv 0$ the matrix $(\Hat V(s_j-s_k))$ is negative definite
for any choice and any number of mutually distinct points $s_j\in \RR^n$
by the Bochner theorem.

The above contructions can be also applied to a class of matrix Hamiltonians.
Namely, consider an operator $H_0$ acting in $L^2(\RR^n)\otimes \CC^d$
whose Fourier symbol in the multiplication by a $d\times d$
Hermitian matrix $H_0(p)$. Then there exist unitary matrices $U(p)$, $p\in\RR^n$,
and real-valued continuous functions $p\mapsto\lambda_1(p)$, \dots, $p\mapsto\lambda_d(p)$
with $\lambda_1(p)\le \dots\lambda_2(p)\le\dots\lambda_d(p)$ such that
\[
H_0(p)=U(p)\diag \big(\lambda_1(p),\dots,\lambda_d(p)\big) U^*(p).
\]
We assume that $\lambda_1(p)$ satsfies the same conditions as the symbol $H_0(p)$ in the scalar case. We will use the same notation;
in particular, $\min \lambda_1(p)=\inf \spec H_0=m$.

A class of such matrix operators is delivered by spin-orbit Hamiltonians \cite{Win}
acting in $L^2(\RR^2)\otimes\CC^2$ and given by the matrices
\begin{equation}
            \label{eq-hap}
H_0(p)=\begin{pmatrix}
p^2 & a(p)\\
\overline{a(p)} & p^2
\end{pmatrix}
\end{equation}
with some linear functions $a$. In particular, the case
$a(p)=\alpha(p_2+ip_1)$ corresponds to the Rashba Hamiltonian~\cite{BR},
and $a(p)=-\alpha(p_1+i p_2)$ gives the Dresselhaus Hamiltonian~\cite{RS};
in both cases $\alpha$ is a non-zero constant. Here one has
$\lambda_1(p)=p^2-\big|a(p)\big|$, and the minimum $-\alpha^2/4$
is attained at the circle $|p|=|\alpha|/2$.

Again consider a \emph{scalar} real-valued potential $V\in L^1(\RR^n)$.
Assume that the operator $H_0+V$ defined through the form sum
is self-adjoint and that $\inf\spec_{\ess} (H_0+V)=\inf\spec_{\ess} H_0=m$.
The preservation of the essential spectrum for the above Rashba and Dresselhaus Hamiltonians
is guaranteed, e.g. for $V\in L^1(\RR^2)\cap L^2(\RR^2)$,
which is achieved by comparison with the two-dimensional Laplacian.

Also in this case we can prove an analogue of corollary \ref{corol3}.

\begin{prop}
Let $V\le 0$ and $V\not\equiv 0$, then the matrix Hamiltonian $H$
has infinitely many eigenvalues below $m$.
\end{prop}

\begin{proof}
The proof follows the construction in the proof of proposition \ref{prop1}.
Consider the vector $h=(1,0,\dots,0)^T\in\CC^d$, and for $\varepsilon>0$ denote
$F^\varepsilon_j(p)=U(p) f^\varepsilon_j(p) h $
with the functions $f^\varepsilon_j$ from \eqref{eq-fj}. 
Then one has
\begin{multline*}
\langle F^\varepsilon_j,(H-m) F^\varepsilon_k\rangle=
\int_{\RR^n} \lambda_1(p)\,\overline{f^\varepsilon_j(p)}f^\varepsilon_k(p) dp\\
+\int_{\RR^n} \int_{\RR^n} \Hat V(p-p')
\langle U(p) h,U(p')h\rangle
\overline{f^\varepsilon_j(p)}  f^\varepsilon_k(p')
\, dp\, dp'.
\end{multline*}
By \eqref{eq-v1} and \eqref{eq-v2}, there holds
\begin{multline*}
\lim_{\varepsilon\to 0}\langle F^\varepsilon_j,(H-m) F^\varepsilon_k\rangle
\\=
\int_S \int_S \Hat V(s-s')
\big\langle U(s) h,U(s')h\big\rangle
\overline{\Psi_1(s)}  \Psi_2(s')
\, \omega(ds)\, \omega(ds').
\end{multline*}
By the arguments of proposition \ref{prop1}, the number of eigenvalues
of $H$ below $m$ is not less than the number of negative eigenvalues of the operator $\UU$
acting on $L^2(S,\omega)$ and given by
\[
\UU f(s) =\int_S \Hat V(s-s')
\big\langle U(s) h,U(s')h\big\rangle f(s) ds'.
\]
Again, by
\begin{multline*}
\int_S \int_S \Big|\Hat V(s-s')\big\langle U(s) h,U(s')h\big\rangle\Big|^2 \omega(ds)\,\omega(ds')\\
\le
\int_S \int_S \big|\Hat V(s-s')\big|^2 \omega(ds)\,\omega(ds')
<\infty, 
\end{multline*}
$\UU$ is a compact operator.
Let us show that all eiganvalues of $\UU$ are negative
(this, like in proposition~\ref{prop2}, will mean that $\UU$ has an infinite number
of negative eigenvalues). For $f\in L^2(S,\omega)$ one has
\begin{multline}
          \label{eq-uu}
\langle f,\UU f\rangle
= \int_S \int_S \Hat V(s-s') \big\langle U(s) h,U(s')h\big\rangle
\overline{f(s)} f(s') \omega(ds) \omega(ds')\\
=
\int _S \int_S \int_\RR \dfrac{V(x) e^{i\langle s'-s,x\rangle}}{(2\pi)^{n/2}} 
\big\langle U(s) h,U(s')h\big\rangle
\overline{f(s)} f(s') dx\,\omega(ds)\, \omega(ds')\\
=\dfrac{1}{(2\pi)^{n/2}}\int_\RR V(x) \big| g(x)\big|^2\,dx\le 0
\end{multline}
with
\[
g(x):=\int_S U(s)f(s)h e^{-i\langle s, x\rangle} \omega(ds).
\]
It remains to show that $0$ is not an eigenvalue of $\UU$.
Assuming $\UU f=0$ we obtain from \eqref{eq-uu} that $g$ vanishes on
the support of $V$ having non-zero Lebesgue measure.
As $g$ is again the Fourier transform on a compactly supported distribution
and hence analytic,
it must vanish everywhere, which means that the vector function $s\mapsto U(s)f(s)h$
is zero a.e. As the matrix $U(s)$ is unitary for any $s$,
this means $f=0$.
\end{proof}
We note that proposition \ref{prop4} and corollary \ref{corol3}
for the Rashba and Dresselhaus Hamiltonians
were shown in in \cite{BGP1} using test functions of a different type.
The above analysis can be extended to the case when the perturbation $V$
is not a potential, but a measure with some regularity conditions.
For the Hamiltonians \eqref{eq-hap} one can still prove
the infiniteness of the discrete spectrum for perturbations by negative measures
supported by curves~\cite{BGP2}.

\bigskip

I thank Grigori Rozenblum for pointing out
some shortcomings in a preliminary version of the paper.
The work was supported in part by the Deutsche Forschungsgemeinschaft.

\end{document}